\documentclass[12pt]{iopart}
\usepackage{graphicx}
\usepackage{units}
\usepackage{hyperref}
\usepackage{iopams}

\usepackage{color}

 \newcommand{\changed}[1]{{#1}}

\begin{document}
\title{\changed{Few-femtosecond resolved imaging of laser-driven nanoplasma expansion}}
\author{C. Peltz$^1$, J.A. Powell$^{2,3,4}$, P. Rupp$^5$, A Summers$^{2,6}$, T. Gorkhover$^{7,8}$, M. Gallei$^9$, I. Halfpap$^10$, E. Antonsson$^10$, B. Langer$^10$, C. Trallero-Herrero$^{2,3}$, C. Graf$^11$, D. Ray$^7$, Q. Liu$^{5,12}$, T. Osipov$^7$, M. Bucher$^7$, K. Ferguson$^7$, S. Möller$^7$, S. Zherebtsov$^{5,7}$, D. Rolles$^2$, E. R\"{u}hl$^10$, G. Coslovich$^7$, R. N. Coffee$^{7,13}$, C. Bostedt$^{14,15,16}$, A. Rudenko $^2$, M.F. Kling $^{5,12}$ and T. Fennel$^{1,17}$}

\address{$^1$ Institute for Physics, Universit\"{a}t Rostock, D-18051 Rostock, Germany}
\address{$^2$ J.R. Macdonald Laboratory, Department of Physics, Kansas State University, Manhattan, Kansas 66506, USA}
\address{$^3$ Department of Physics, University of Connecticut, Storrs, Connecticut 06269, USA}
\address{$^4$ INRS, Énergie, Matériaux et Télécommunications, 1650 Blvd. Lionel Boulet, Varennes, Québec, J3X 1S2, Canada}
\address{$^5$ Department of Physics, Ludwig-Maximilians-Universit\"at Munich, D-85748 Garching, Germany}
\address{$^6$ ICFO - The Institute of Photonic Sciences, The Barcelona Institute of Science and Technology, 08860 Castelldefels, Barcelona, Spain}
\address{$^7$ Linac Coherent Light Source, SLAC National Accelerator Laboratory, Menlo Park, CA 94025, USA}
\address{$^8$ University of Hamburg, Institute for Experimental Physics, 22761 Hamburg, Germany}
\address{$^9$ Saarland University, D-66123 Saarbrücken, Germany}
\address{$^10$ Physical Chemistry, Freie Universit\"{a}t Berlin, D-14195 Berlin, Germany}
\address{$^11$ Department of Chemistry and Biotechnology, Darmstadt University of Applied Sciences, D-64295 Darmstadt, Germany}
\address{$^12$ Max Planck Institute of Quantum Optics, D-85748 Garching, Germany}
\address{$^13$ The Pulse Institute, SLAC National Accelerator Laboratory, Menlo Park, CA 94028, USA}
\address{$^14$ Chemical Sciences and Engineering Division, Argonne National Laboratory, Argonne, Illinois 60439, USA}
\address{$^15$ Paul-Scherrer Institute, CH-5232 Villigen PSI, Switzerland}
\address{$^16$ LUXS Laboratory for Ultrafast X-ray Sciences, Institute of Chemical Sciences and Engineering, \'Ecole Polytechnique F\'ed\'erale de Lausanne (EPFL), CH-1015 Lausanne, Switzerland}
\address{$^17$ Max Born Institute, D-12489 Berlin, Germany}

\ead{thomas.fennel@uni-rostock.de}

\date{\today}
\begin{abstract}
{The free expansion of a planar plasma surface is a fundamental non-equilibrium process relevant for various fields but as-yet experimentally still difficult to capture. The significance of the associated spatiotemporal plasma motion ranges from astrophysics and controlled fusion to laser machining, surface high-harmonic generation, plasma mirrors, and laser-driven particle acceleration. Here, we show that x-ray coherent diffractive imaging can surpass  existing approaches and enables the quantitative real-time analysis of the sudden free expansion of laser-heated nanoplasmas. For laser-ionized SiO$_2$ nanospheres, we resolve the formation of the emerging nearly self-similar plasma profile evolution and expose the so far inaccessible shell-wise expansion dynamics including the associated startup delay and rarefaction front velocity. Our results establish time-resolved diffractive imaging as an accurate quantitative diagnostic platform for tracing and characterizing plasma expansion and indicate the possibility to resolve various laser-driven processes including shock formation and wave-breaking phenomena with unprecedented resolution.} 
\end{abstract}
\noindent{\it Keywords\/}: coherent diffractive imaging, plasma expansion, laser nanoparticle interaction

\submitto{\NJP}
\maketitle

\section{Introduction}
The sudden free expansion of a plasma with an initially sharp density step is a key model process for astrophysical flows
\cite{BorovskyAstrophys280_802_1984}, fusion pellet ablation~\cite{ChangNuclFus1980,MiloraNuclFus1995}, and plasma dynamics initiated by discharges~\cite{TanbergPR1930} or wire explosions~\cite{EiseltZPhys1952,NashPhysFluids1961}. Especially the occurrence and structure of expansion modes with self-similar density  evolution~\cite{GurevichJETP1966} has been subject of vivid debate and intense analytical and numerical analysis~\cite{True_PhysFluids1981,SackPR156_311_1987,MurakamiPoP2006,BeckPPCF2009}. The significance of free plasma expansion culminates in the realm of intense laser-matter interactions at surfaces due to its central role for laser-driven monoenergetic ion beam generation~\cite{MoraPRL2003,DaidoRoP2012}, vacuum acceleration of electrons~\cite{ThevenetNatPhys2016}, high-harmonic generation through relativistic oscillating mirrors~\cite{BulanovPhysPlasmas1994,MourouRMP2006,RoedelPRL2012} or coherent wake emission~\cite{QuerePRL2006,Borot_NatPhys2012}, and pulse cleaning via transient plasma mirrors~\cite{ThauryNatPhys2007}. In all these cases, the plasma density profile at the surface is a pivotal element or critical control parameter for the main process. Moreover, access to the surface expansion speed of isochorically heated plasmas enables characterization of thermophysical properties of warm dense matter~\cite{Bang_SciRep5_14318(2015),Bang_SciRep6_29441(2016)}, underlining the far-reaching implications of an accurate characterization and understanding of plasma profile and expansion velocity dynamics.

Established approaches for the experimental charactization of laser-induced plasma expansion include shadowgraphy and fluorescence imaging~\cite{Bang_SciRep5_14318(2015),HarilalPOP2017}, optical interferometry~\cite{Wang.PhysRevB.92.174114(2015)}, Doppler spectroscopy~\cite{MondalPRL2010}, and proton imaging~\cite{RomagnaniPRL2005}, each with its  specific merits. However, for the complete characterization of the microscopic expansion process including the formation and evolution of self-similar expansion modes, simultaneous resolution of density and velocity profiles on nanometer spatial and femtosecond temporal scales or even beyond is essential, which exceeds current capabilities  and is so far restricted to simulations.

\begin{figure*}[t]
	\includegraphics[width=1.0\textwidth]{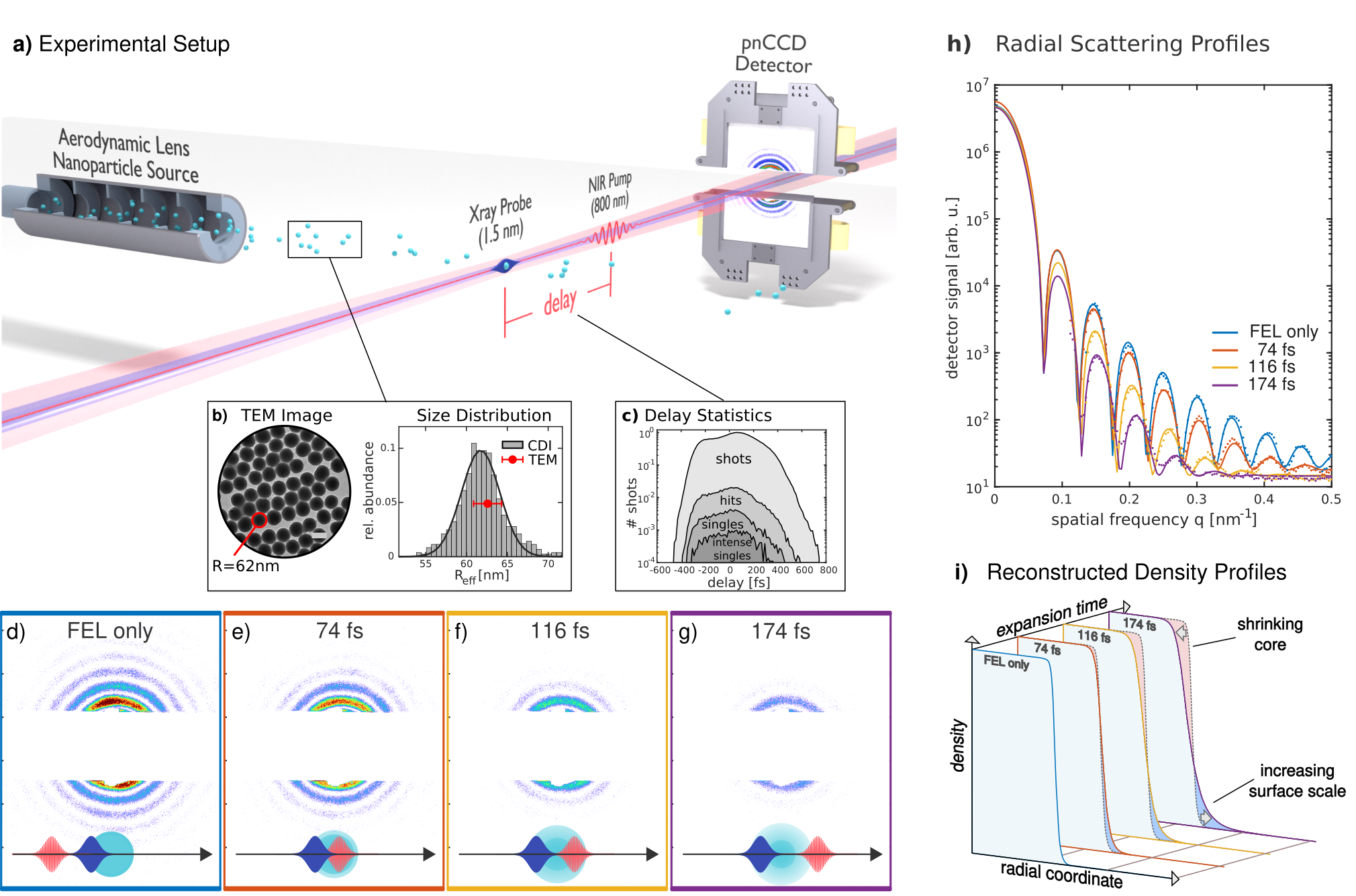}
	\caption{ {\bf Sketch of the experiment.} (a) A beam of chemically prepared SiO$_2$ nanospheres with 62\,nm mean radius was delivered into the interaction region by an aerosol source employing aerodynamic lens focusing. The width of the size distribution is characterized using both TEM images and the x-ray scattering pattern [see panel (b)]. X-ray pulses with 800~eV photon energy, 40\,fs duration, and 1~mJ pulse energy are delivered at 120~Hz repetition rate and  were focussed into the interaction region using a Kirkpatrick–Baez mirror system. They are merged with synchronized 50~fs NIR laser pulses at 800~nm central wavelength and intensity \unit{8$\times$10\textsuperscript{14}}{cm\textsuperscript{-2}} using a holey mirror placed behind the x-ray focusing optics. Scattering images are recorded by two pnCCD detector panels placed 40\,cm behind the interaction zone. The gap between the detector panels transmits the primary beams of the x-ray and NIR lasers. Scattering images recorded for each laser shot are classified as hits, singles, and intense singles [see histograms in (c)] and only the latter are used for further analysis. The delay dependent scattering images are fitted with a model (see (d-g) for representative measured patterns and (h) for lineouts and associated fit results). (i) The associated reconstructed density profiles are displayed as function of the effective radial coordinate $r_{\rm eff}=\kappa+R_{\rm eff}$ (see text) and encode the plasma expansion. The profile dynamics resembles the laser-induced core shrinking and surface layer expansion (as indicated). \label{fig_setup}}
\end{figure*}

An emerging alternative towards ultrafast and accurate tracing of plasma dynamics is time-resolved coherent diffractive imaging of laser-driven free nanoparticles using x-ray free electron lasers (XFEL)~\cite{Chapman2007}. Complete atomistic pump-probe simulations of ultrafast near-infrared (NIR) laser-driven ionization and heating of hydrogen nanospheres have predicted the rapid formation of a nearly self-similar expansion mode for the initially step-like density profile~\cite{PeltzPRL2014}. The main short-term dynamics is predicted to unfold within few femtoseconds and includes an inward-motion of the rarefaction front expressed by a decreasing radius of the dense plasma core, an increasing softness of the density edge, and the spreading of the expanding surface layer that maintains a density profile with an asymptotically exponential radial decay. Most importantly, a direct mapping between x-ray scattering patterns and the associated plasma profile parameters has been identified for nearly spherical particles, promising the experimental characterization of the density profile evolution with high spatial and temporal resolution. 

In a pioneering study, Gorkhover \emph{et al.} demonstrated a substantial expansion of the surface layer of laser-heated Xe clusters on the few hundred fs time scale~\cite{GorkhoverNatPhoton2016} by analyzing the evolution of the angular decay of the ring-shaped diffraction fringes. Wide-angle Bragg scattering with hard XFEL pulses can yield insights into the dynamics of strongly pumped nanoplasmas~\cite{Ferguson2016}, indicating that the loss of order proceeds from the surface to the inner core of the cluster~\cite{Nishiyama2019}. \changed{These results reflect a natural extension of previous successful studies on ultrafast structural changes associated with non-thermal heating and nonequilibrium phase transitions via time-resolved Bragg scattering at surfaces~\cite{Rousse_Nature_2001,Lindenberg_Science_2005}.} Diffraction experiments on laser-driven metal gratings further support high sensitivity of scattering patterns to the ultrafast  density profile evolution at the surface~\cite{KlugePRX2018}.
However, the envisaged complete tracing of the laser-induced plasma expansion has so far remained a challenge as it requires the combination of (i) accurate knowledge of the target's initial size and shape, (ii) an unambigous profile reconstruction based on an appropriate few-parameter profile model, (iii) precise single-shot pulse timing to remedy the otherwise pathological time jitter of free-electron laser sources when operated in self-amplified spontaneous emission (SASE) mode, and (iv) a sufficently large dataset to  trace the profile evolution systematically. 

Here, we report simultaneous realization of all the above prerequisites and demonstrate quantitative time-resolved imaging of laser-induced nanoplasma expansion. The NIR pump x-ray diffraction probe experiment was performed for isolated SiO$_2$ nanospheres at the Linac Coherent Light Source (LCLS) at SLAC and exposes the density profile evolution with unprecedented temporal and spatial resolution. Moreover, our analysis provides access to the so far inaccessible shell-resolved expansion startup delay and velocity buildup as well as the  associated rarefaction front velocity. These results define not only a critcal benchmark for theory but also enable the systematic study of the impact of laser characteristics and material properties on the ultrafast formation and relaxation of laser-driven nanoplasmas.

\section{Experiment}

The measurement was performed at the AMO hutch of LCLS using the LAMP multipurpose endstation, see schematic setup in Fig.~\ref{fig_setup}(a). The beam of SiO$_2$ nanospheres with narrow distribution of size (\unit[62]{nm} mean radius, see Appendix~\ref{Appendix:target_preparation}) and deformation crossed the optical axis of the collinear NIR and x-ray pulses. For the employed \unit[50]{fs} NIR pulses at \unit[800]{nm} central wavelength and peak intensity \unit{8$\times$10\textsuperscript{14}}{Wcm\textsuperscript{-2}} (see Appendix~\ref{Appendix:intensity_calibration}), strong and nearly homogeneous ionization of the resulting nanoplasma to atomic charge states up to $q\approx 6$ is expected~\cite{PeltzPRL2014,VarinPRL2012}. 
The single-shots scattering images resulting from the temporally synchronized illumination of the particles with \unit[40]{fs} x-ray pulses at \unit[800]{eV} photon energy are recorded with two rectangular pnCCD detector panels~\cite{Strueder2010} that were covered by a spectral filter (see Appendix~\ref{Appendix:scattering_detector_spectral_filters}) to suppress NIR photons and XUV fluorescence from the heated nanoplasma. Employing a smaller x-ray spot size and selecting only bright diffraction images enables the exclusive analysis of particle scattering images from a region that was homogeneously pumped by the NIR laser.  Downstream the optical axis, the accurate relative timing of the x-ray and NIR pulses is measured on a single shot basis with few-femtosecond accuracy (see Appendix~\ref{Appendix:laser_and_timing})~\cite{Schorb:10,Schlotter:10,Harmand13}.

The single-shot diffraction images were systematically recorded as function of pulse delay.  While scattering patterns with bright, nearly spherical fringes from unpumped particles are observed for negative pulse delay (x-rays first), the fringe features begin to move and fade with increasing delay for pumped particles (positive delays), see Fig.~\ref{fig_setup}(d-g). The evolution of the fringe pattern represents our main observable and encodes the plasma density dynamics. Note that images from unpumped particles enable the in-situ characterization of the initial target size distribution (see Fig.~\ref{fig_setup}b).

\section{Analysis}
\subsection{Shot Statistics}
From $1.3\times10^6$ recorded scattering images we found 24500 clearly identifiable hits. After removing faint images, multiparticle hits that display Newton rings~\cite{BostedtJPB2010,RuppNJP2012}, and hits from clustered nanoparticles, a dataset of 5700 bright single-shot images with clear fringe structures has been systematically analyzed as described below. Out of that, a subset of 1070 images was assigned to the high intensity region of the x-ray focus which is associated with homogeneous NIR pumping. It should be emphasized that the latter discrimination step cannot be made based on the scatting pattern alone but requires the quantitative fit of the image with our model to determine the associated incident x-ray fluence (see Appendix~\ref{Appendix:fitting_procedure}). The distributions of classified images versus pulse delay are displayed in Fig.~\ref{fig_setup}c.

\subsection{Individual Image Analysis}
\begin{figure}
	\centering
	\includegraphics[width=0.6\textwidth]{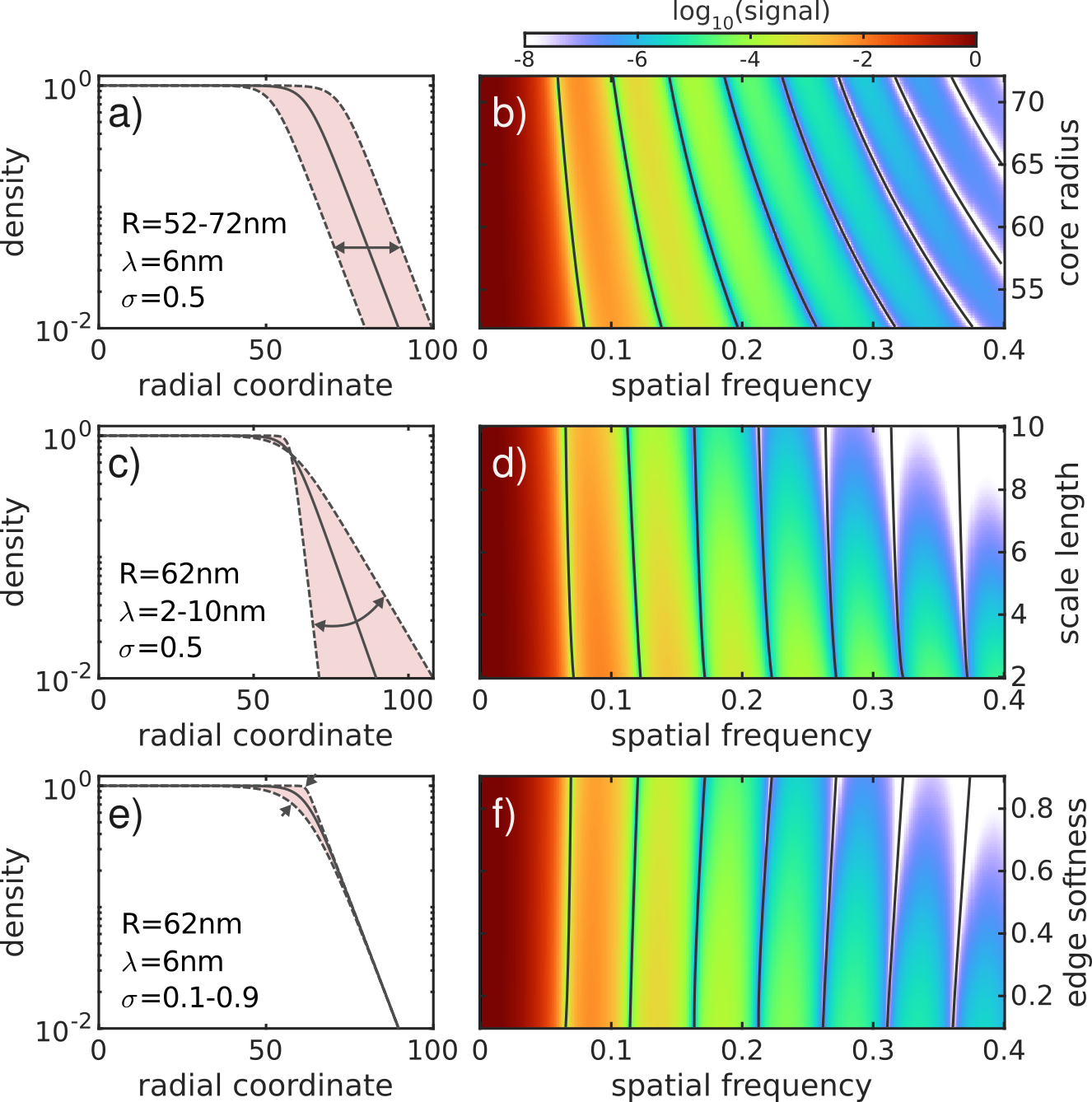}
	\caption{Systematic analysis of the correlation between the density profile parameters core radius, scale length and edge softness and the central signatures of the scattering pattern, i.e. fringe spacing and signal slope, evaluated in the limit of spherical particle symmetry. Panels (a,c,e) depict the employed density profile reference as a solid black line and the effect of the selective variation of a single profile parameter as light red areas. Corresponding line-outs of the resulting scattering pattern are shown in panels (b,d,f). \label{fig_dens_vs_scatt}}
\end{figure}

In the small-angle scattering limit, the far-field  scattering pattern encode a two-dimensional projection of the particle density. Nevertheless, for three-dimensional targets with spherical symmetry, the unique and unrestricted reconstruction of the radial density profile would still be possible if the complete scattering image can be recorded. In real-world experiments, however, a substantial part of information is lost due to the unavoidable detector gaps in pump-probe experiments and limited signal at larger scattering angles. For the quantitative reconstruction of the density profile from the available information contained in the single-shot scattering images an appropriate shape model is imperative. In our analysis we use a simulation-guided few-parameter density profile~\cite{PeltzPRL2014} 
and retrieve the profile parameters by systematically  fitting simulated scattering patterns to the individual measured scattering images. The employed shape model is based on an electron density profile described by a modified Fermi function with
\begin{equation}
n(\vec{r})= \frac{n_0}{\left[\exp\left(\frac{\kappa(\vec{r})}{  \lambda \sigma}\right)+1\right]^\sigma}
\label{eq_profile}.
\end{equation}
Here $n_0$ is the core electron density, $\kappa$ is the local surface offset of the sampling point $\vec{r}$, $\sigma$ is the edge softness parameter, and $\lambda$ is the scale length of the density decay. A sharp density edge corresponds to the limit $\sigma\rightarrow0$. For a spherical system with a reference surface at radius $R$, the surface offset is $\kappa=r-R$. The variation of radius, scale length, and softness are connected with specific signatures in the scattering pattern, as illustrated in Fig.~\ref{fig_dens_vs_scatt} for a spherical system. For a fixed softness parameter, the radius and scale length are almost independently coupled to the fringe spacing and angular decay of the radial scattering profile (cf. Figs.~\ref{fig_dens_vs_scatt} a-d), respectively. As the underlying  mapping between these profile parameters and the signatures in the scattering images is a function of the softness, the latter must be known or has to be determined independently for a unique parameter reconstruction. This requirement is illustrated in Fig.~\ref{fig_dens_vs_scatt}e-f, where the variation of the edge softness is shown to also modify the fringe spacing and the angular decay of the scattering profile. It should be noted that in previous work, the problem was usually circumvented by assuming an unphysical sharp edge ($\sigma\rightarrow0$). In our analysis we determine the finite edge softness using an additional constraint, as described in detail below.  

In the description of the scattering process, both plasma electrons as well as atomically bound electrons with binding energies smaller than the X-ray photon energy can be treated as free electrons. In contrast, the scattering contribution of strongly bound electrons is negligible, such as in the case of the silicon K-shell electrons in our scenario. It should further be noted, that the K-shell electrons of Si are effectively unreachable for the NIR-driven ionization. Hence the Si K-shell can be neglected for both ionization and the scattering process. As a result and assuming that the fraction of emitted electrons is negligible for the large particles considered here, the density of actively scattering electrons is representative for the atomic density and their number can be assumed to be constant. 

In our analysis the slight ellipsoidal deformation of the nanospheres resulting from the particle synthesis must be included for high accuracy fits (see Appendix~\ref{Appendix:small_angle_scattering}). For the resulting ellipsoidal reference surface with direction dependent radius, the local surface offset is defined as  $\kappa=\left[\vec{r}-\vec{\rho}\,\right] \cdot \vec{n}$, where $\vec{\rho}(\vec{r})$ specifies the closest point on the reference surface and $\vec{n}$ is the associated unit vector of the surface normal. The neglect of directional variations of the remaining relevant parameters $\sigma$ and $\lambda$ is justified by the small anisotropy predicted by microscopic calculations \cite{PeltzPRL2014}. 

For a given value of the edge softness parameter $\sigma$, high quality fits of individual images can be performed and yield the scale length $\lambda$ and the minimal and maximal core radii  $\rho_{\rm{min}}$ and $\rho_{\rm{max}}$ that characterize the minor and major axis of the projected density associated with the ellipsoidal control surface. For further analysis we use the effective radius $R_\textrm{eff}=\frac{\rho_{\rm{min}}+\rho_{\rm{max}}}{2}$. 
So far, however, the  remaining dependence of the scale length and core radius on the edge softness results in an apparent critical ambiguity for the analysis. We resolve this ambiguity by exploiting the conservation of the number of scattering electrons (i.e conservation of mass in the case of charge neutrality), as described in the following.

\subsection{Determining the edge softness evolution}
\begin{figure*}[t]
	\includegraphics[width=1.0\textwidth]{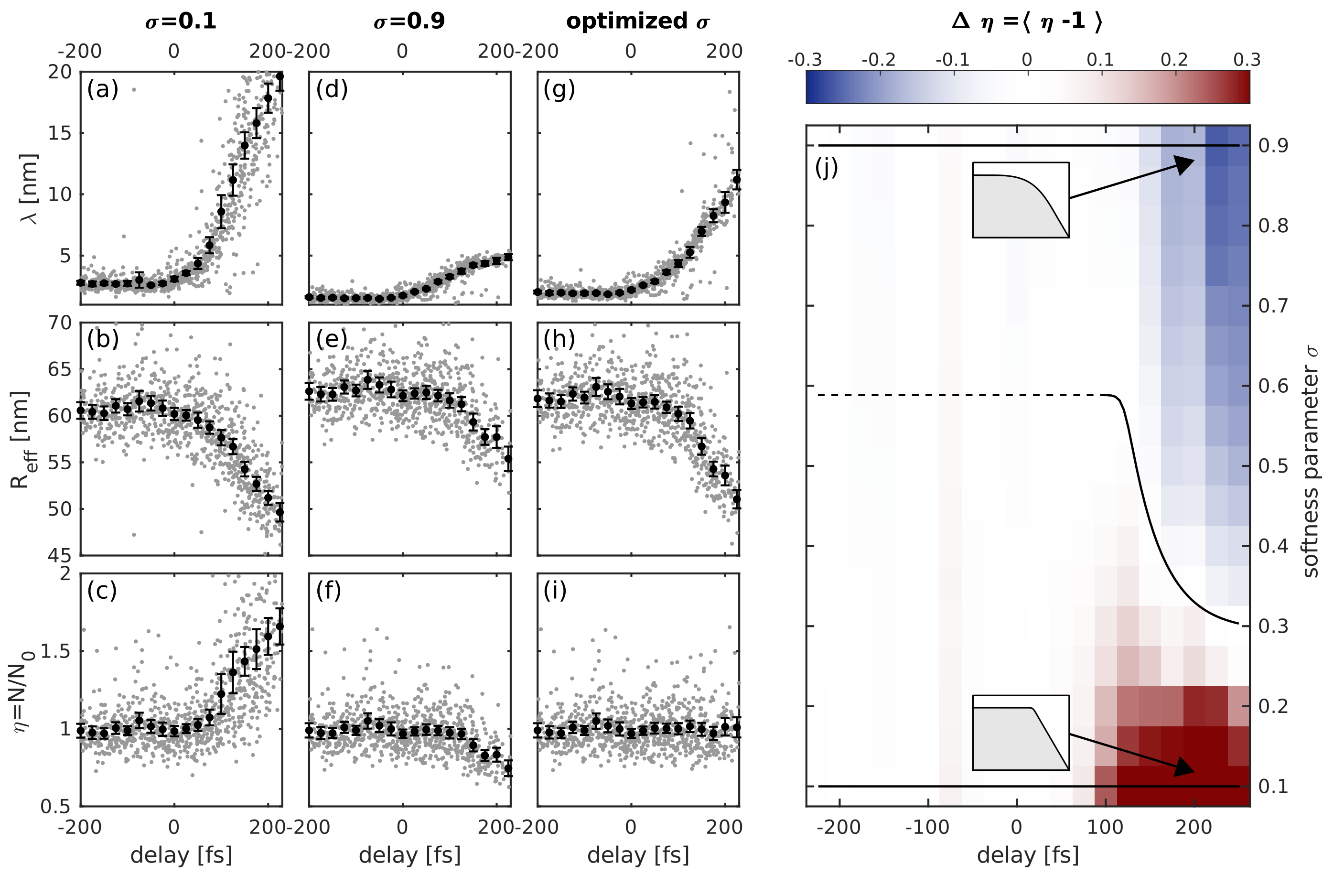}
	\caption{{\bf Reconstruction of the density profile parameters.} (a-i) Temporal evolution of surface width $\lambda$, effective core radius  $R_{\rm eff}$ and associated relative mass $\eta=N/N_0$ reflecting the particle number $N=\int n(\vec{r})d^3 r$ ($\propto$ total mass) normalized to the average value for unpumped shots $N_0$. Evolutions are displayed assuming a small (a-c), large (d-f), and the optimized softness parameter (g-i). Each gray dot represents an individual scattering image while the black symbols indicate the time-binned mean values and their corresponding standard deviations. A fixed sharp edge ($\sigma=0.1$) leads to a strong overestimation of the particle number at large delays by roughly 70\% (a-c). A fixed soft edge $\sigma=0.9$ leads to the opposite effect, i.e. an unphysical mass loss with delay (d-f). A systematic analysis of the mass mismatch evolution versus softness yields a unique optimal path in the delay-softness plane that reflects mass conservation (j). Using  $\sigma(t)$ determined by this optimal path leads to the final and physically consistent parameter evolution (g-j).} 
	\label{fig2}
\end{figure*}

The treatment of the scattering problem via the few-parameter profile enables systematic forward fitting of a large dataset with thousands of scattering images. Figure~\ref{fig2}a,b displays fit results for the delay dependent evolution of  scale length and effective radius when assuming a fixed very sharp edge ($\sigma=0.1$). In this case, the associated total mass (number of electrons contributing to the scattering as given by the integral of the electron density) would increase by almost 90\% during the first $200\,\rm{fs}$ of the expansion (Fig.~\ref{fig2}c), which is obviously unphysical. The profile parameters retrieved assuming a very soft edge ($\sigma=0.9$) differ substantially (Fig.~\ref{fig2}d,e) and, in turn, would correspond to a mass decrease by more than 20\% (Fig.~\ref{fig2}f). The mass mismatch evolution map resulting from fits with systematically varied softness parameters is shown in false color representation in  Fig.~\ref{fig2}j, where the above discussed representative cases are indicated by horizontal lines (top and bottom). While the impact of the edge softness on the mismatch is little for unpumped particles as its effect vanishes in the limit of small scale length (sharp density step), the map shows that mass conservation during expansion requires a dynamical change of the softness parameter, expressed by the \emph{path of mass conservation} (white color) in Fig.~\ref{fig2}j. The delay dependent softness associated with conserved total mass  (black curve) defines the final and physically consistent evolution of all parameters and their mean values, i.e. softness, scale length and core radius, see Figs.~\ref{fig2}g-j. Only these parameters have been used for further analysis and in Fig.~\ref{fig_setup}.

\begin{figure*}[t]
	\includegraphics[width=1.0\textwidth]{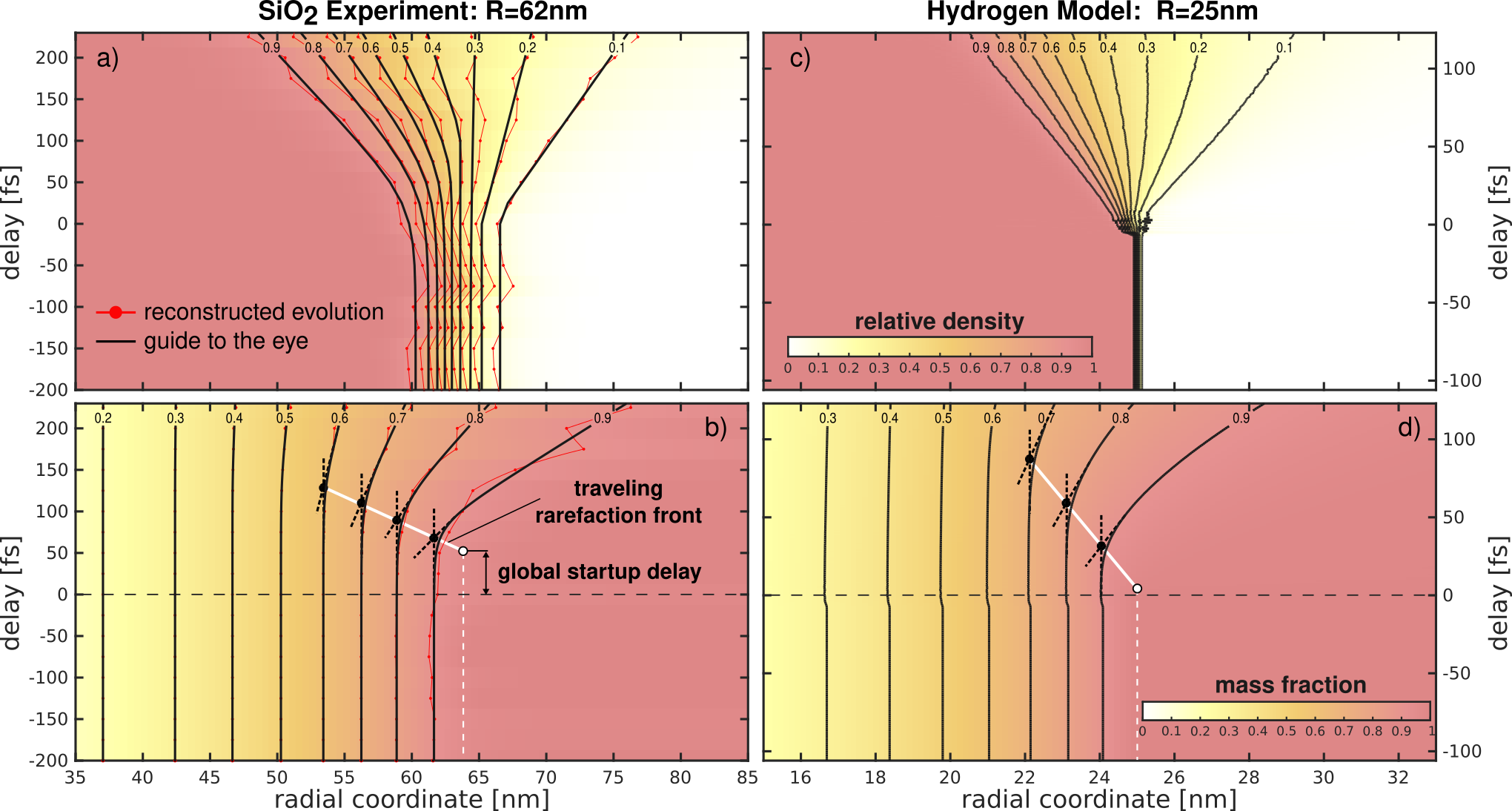}
\caption{{\bf Evolution of radial density and mass fraction profiles.} (a,b) Reconstructed spatio-temporal density map and corresponding mass fraction map (color coded) with selected contour levels (red, levels as indicated) together with fit curves (solid) as guides to the eye. For qualitative comparison of the experimental features with available plasma model calculations, panels (c,d) show density and mass fraction maps (contour levels as indicated) as extracted from simulation results for hydrogen spheres~\cite{PeltzPRL2014}. Shell-specific expansion start-up times (black dots) in (b,d) are estimated as the intersection of the tangent to the trajectory at 2.5$\%$ relative shell expansion with the initial shell radius (c.f. black dotted lines). The global start-up delay (circle) is determined from extrapolation to the initial surface (see text), which we define as the radius where the initial relative density profile has dropped to $1/e$ (white dotted line). White solid lines indicate the motion of the rarefaction front. \label{fig3}}
\end{figure*}

\section{Results and discussion}
The resulting time-dependent density map in Fig.~\ref{fig3}a displays the retrieved mean density profile evolution and reflects the core depletion and ablation dynamics of the nanoplasma. Both the decreasing radius of the remaining high-density core as well as the increasing width of the surface layer can be clearly identified and tracked in time, see the evolution of density isolines. Note that the initial finite width of the density step in the experimental data for unpumped particles (negative delays) is attributed to surface roughness and deviations from an ellipsoidal shape. Minimizing the initial roughness and surface porosity via advanced target preparation techniques is expected to remedy this effect~\cite{RaschpichlerJPCC20}. Most importantly, the continuous sampling of the mean density profile along the time axis further allows to track the expansion startup and velocity buildup for individual shells - an information that could not be accessed with existing plasma imaging techniques. To that end we analyze the evolution of the normalized radial density integral. The resulting mass fraction map (Fig.~\ref{fig3}b) reflects the relative particle mass enclosed in a given radius as function of time. Contours of equal mass fraction (see lines), in turn, reflect the expansion of individual shells and define meaningful shell-specific trajectories if overrun effects (i.e. if inner shells overrun outer ones) can be excluded. \changed{Note that the employed geometry model is ideally suited to describe the expansion dynamics after laser excitation. It is not constructed to capture e.g. the sub-cycle dynamics during laser excitation and thus describes an averaged density dynamics during the excitation.}

The evolution of adjacent shell trajectories shows that the surface layer expansion does not proceed with a single speed. From the shell-resolved trajectories we can extract the specific buildup and saturation of the respective shell velocities as well as the associated start-up delay. A comparison of the shell-specific start-up delays (symbols in Fig.~\ref{fig3}b), however, shows an almost constant velocity of the inward-traveling rarefaction front with radial speed $v_{\rm front}^{\rm exp}=-0.135\,{\rm nm/fs}$ and defines a global startup delay of $\tau_{\rm startup}^{\rm exp}=52\,$fs via interpolation to the initial surface radius (see Fig.~\ref{fig3}b). \changed{The global startup delay has been determined only from the dynamics outside the NIR-excitation window and is therefore independent of the sub-cycle dynamics during laser excitation. It does not necessarily specify the actual beginning of the plasma motion, but reflects a meaningful parameter to characterize the rarefaction wave. Both} central parameters, i.e. the front velocity and the global start-up delay, are assumed to be characteristic observables for the dependence of the plasma formation and relaxation dynamics on laser parameters and material properties and are thus of particular interest for comparison with theory.

With existing models, a full microscopic modelling of the expansion dynamics for the experimental scenario is still out of reach. However, to evaluate the significance and generic character of the observed features, we perform a qualitative comparison with the published microscopic particle-in-cell simulations for hydrogen spheres~\cite{PeltzPRL2014}, see Fig.~\ref{fig3}c,d. The simulation results are displayed for axis limits that correspond to similar relative changes of parameters as in the experiment. Though absolute spatial and temporal scales are different, the general structure of the density evolution and shell-resolved expansion trajectories shows excellent agreement. In particular, a finite start-up delay 
of $\tau_{\rm startup}^{\rm sim}=7\,$fs and a nearly constant rarefaction front velocity of $v_{\rm front}^{\rm exp}=-0.035\,{\rm nm/fs}$ are predicted by the simulations, supporting  general applicability of our analysis and the significance of these key parameters. 
In particular, our results demonstrate the accessibility of these parameters in time-resolved diffraction studies, highlighting their potential for the quantitative characterization of both laser-driven plasma formation and the resulting relaxation in both experiment and theory. The development of theoretical models that can explain the impact of  material and laser parameters is expected to allow the description and optimization of such processes for arbitrary target shapes and laser configurations. 

\section{Conclusions}
Our time-resolved diffraction experiment has revealed the quantitative density profile evolution emerging in the early stage of sudden laser-induced nanoplasma expansion for silica nanospheres. The work exposes the so-far inaccessible onset, buildup, and stabilization of the shell-resolved expansion velocity. The observed rarefaction front velocity is comparable with the disorder front velocity reported in~\cite{Nishiyama2019,Niozu_PRX_21}, indicating a potential connection between the two observables that remains to be clarified. The current results provide a critical reference for theory and underpin the unprecedented potential of diffractive imaging for the precise characterization of ultrafast laser-plasma processes, with far-reaching implications for both fundamental research on 
laser-induced self-similar plasma expansion and applications of intense laser-matter interactions at plasma surfaces in general. The feasibility of ultrafast imaging with simultaneous high spatial and temporal resolution~\changed{\cite{Duris_NatPho_2020}} is expected to also enable tracking of other ultrafast plasma dynamics, such as bunching and shock formation in the expansion process~\cite{SackPR156_311_1987} or nonlinear plasma wave formation and breaking~\cite{VarinPRL2012}. \changed{Such experiments,
however, require an appropriately adjusted geometry model to capture and parametrize the relevant features, e.g. to describe density jumps in case of shocks or regular density modulations in case of wave phenomena.} Our method is further anticipated to be transferable to thin cylindrical or even flat jets, opening routes to the systematic tracking of plasma formation, wave dynamics, and the initial phases of ablation in various materials including biologically relevant liquids such as liquid water.  

\ack
Use of the Linac Coherent Light Source (LCLS), SLAC National Accelerator Laboratory, is supported by the U.S. Department of Energy, Office of Science, Office of Basic Energy Sciences under Contract No. DE-AC02-76SF00515. RNC and GC are funded through the LCLS under contract number DE-AC02-76SF00515. This material is based on work supported by the U.S. Department of Energy, Office of Basic Energy Sciences, Division of Chemical Sciences, Geosciences, and Biosciences through Argonne National Laboratory under contract DE-AC02-06CH11357. A.R. and D.R. were supported by the US Department of Energy, Office of Science, Office of Basic Energy Sciences, Division of Chemical, Geological and Biological Sciences (contract no. DE-FG02-86ER13491). J.A.P and C.T.-H. acknowledge support from the Air Force Office of Scientific Research (contract no. FA9550-17-1-0369). We thank Al Rankin for his help in preparing the nanoparticle injector for this experiment and Julia Kredel for support in the particle synthesis. \changed{C.P. and T.F acknowledge support by the Deutsche Forschungsgemeinschaft (DFG, German Research Foundation) - SFB 1477 "Light-Matter Interactions at Interfaces", project number 441234705. T.F. acknowldedges financial support from the Deutsche Forschungsgemeinschaft within the Heisenberg programme (IDs: 315210756, 398382624, 436382461) and via DFG Priority Programme QUTIF (ID: 281272685)}. Computer time was provided by the North-German Supercomputing Alliance (HLRN) via project mvp00013. M.F.K. acknowledges support from the Max Planck Society via the Max Planck Fellow program, and the DFG via the Priority Programme QUTIF (ID: 281272685). S.Z. acknowledges support from the DFG via project no. 322422731. C.B. acknowledges support via the Swiss National Science Foundation National Center of Competence in Research – Molecular Ultrafast Science and Technology NCCR - MUST. 

\appendix
\section*{Appendix}
\section{Target preparation and characterization}
\label{Appendix:target_preparation}
\begin{figure}
    \centering
	\includegraphics[width=0.65\textwidth]{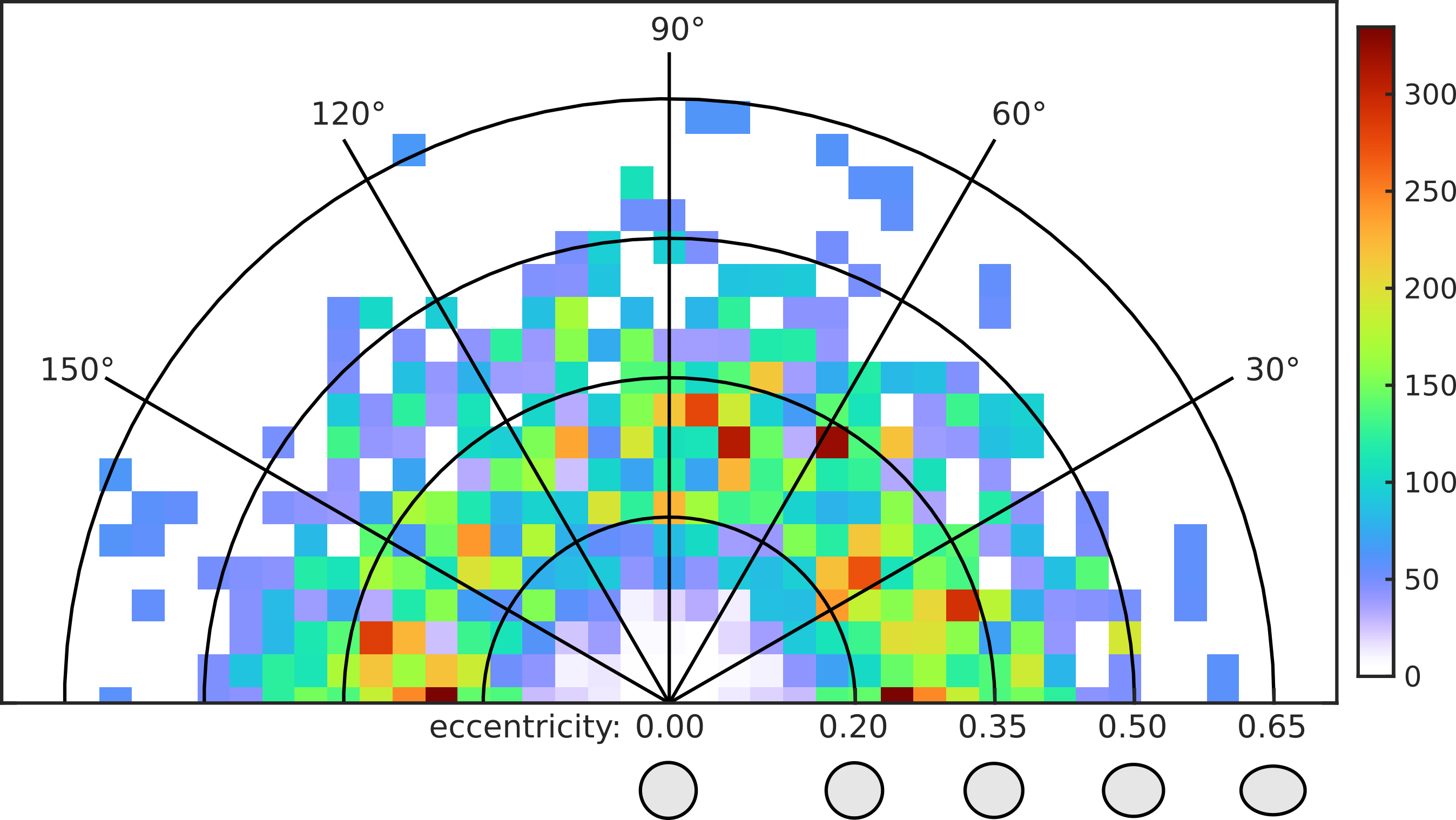}
	\caption{\changed{Distribution of eccentricity and orientation for individual unpumped particles as determined from the scattering images.}
		\label{fig_eccentricit_distribution}}
\end{figure}

Silica nanoparticles were produced by the St\"ober method~\cite{STOBER68}. First, small seed nanoparticles were prepared by adding \unit[21]{g} of TEOS, \unit[28]{mL} of ammonia solution (25\%wt. in water) and \unit[1]{mL} of water to \unit[530]{mL} of ethanol and stirring the mixture for \unit[24]{h}. A further shell was grown on the seed nanoparticles via the seeded growth method~\cite{Graf2002} until the desired particle size was reached. All samples were stored in ultra-pure ethanol after cleaning. The used particle batch was characterized by means of transmission electron microscopy (TEM) as well as by fits of the scattering pattern from unpumped particles (cf. Fig.~\ref{fig_setup}b). The TEM analysis yields an average diameter of $125.2$\,nm and a polydispersity of 1.8$\%$ while a mean diameter of $123.5\,$nm and a polydispersity of 2.3$\%$ is determined from the scattering pattern. Note that these values are consistent when considering the estimated systematic error of ca. 2$\%$ in the analysis of the diffraction data due to uncertainty of the exact distance of the active detector plane from the interaction point and the fluctuation of the central X-ray wavelength. \changed{Further, the systematic analysis of the scattering pattern reveals an average eccentricity $e=\sqrt{1-(\frac{\rho_{\rm{min}}}{\rho_{\rm{max}}})^2}$ of around 0.3 for unpumped particles (cf. Fig.~\ref{fig_eccentricit_distribution}). The associated differences in the ring positions along the minor and major axes in the scattering image would result in significant overlap of neighboring maxima and minima when performing circular averaging. The high accuracy reproduction of the scattering images necessary for our forward fitting approach can therefore only be achieved when the elliptic nature of the particles is taken directly into account (see \ref{Appendix:small_angle_scattering} for more details on the forward fitting method).} Suitable aliquots of the mother suspension were dispersed in larger volumes in deionized water at a concentration of \unit[1]{g/l}, then briefly sonicated, and used in the experiments. The nanoparticles are aerosolized with nitrogen. The aerosol was subsequently dried in a counterflow membrane dryer employing dry nitrogen, introduced into the vacuum chamber through a $120\,\mathrm{\mu m}$ glass nozzle, and collimated by a differentially pumped aerodynamic lens to form a beam of nanoparticles.

\section{IR-Intensity Calibration}
\label{Appendix:intensity_calibration}
\begin{figure}
    \centering
	\includegraphics[width=0.6\textwidth]{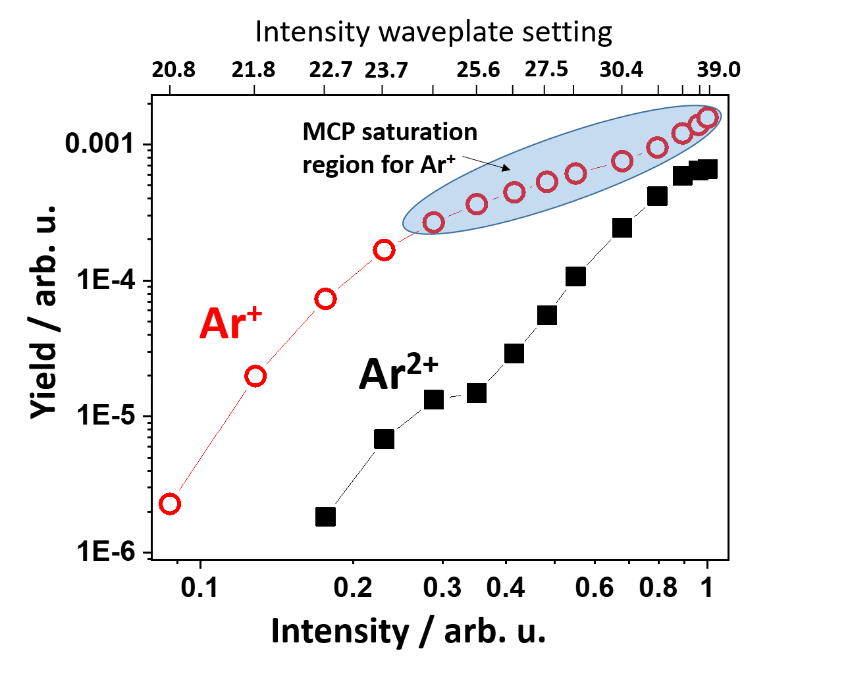}
	\caption{Measured yields of singly and doubly charged Ar ions used for NIR intensity calibration. The top axis shows the settings of the $\lambda$/2 wave plate used to control the intensity, whereas the bottom axis indicates the corresponding transmission, i.e., the relative NIR pulse intensity. For the wave plate settings above 23.7, Ar\textsuperscript{+} signal is significantly affected by the detector saturation.
		\label{fig_int_calibration}}
\end{figure}

Because neither the NIR pulse energy on target (after the holey mirror) nor its focus profile was directly measured, the absolute calibration of the peak intensity of the NIR pulses was performed based on the intensity-dependent yields of Ar\textsuperscript{+} and Ar\textsuperscript{2+} ions recorded using an ion time-of-flight spectrometer. The intensity of the NIR pulse was controlled using a $\lambda$/2 wave plate set in front of a polarizing beam splitter. The measured Ar\textsuperscript{+} and Ar\textsuperscript{2+} ion yields are shown in Fig.~\ref{fig_int_calibration} as a function of the wave plate settings. Since the detector signal for Ar\textsuperscript{+} was saturated in the high-intensity region, we used the intensity-dependent Ar\textsuperscript{2+} signal for the calibration. By matching the Ar\textsuperscript{2+} yield curve to the literature data~\cite{Larochelle1998,Guo1998}, we estimate that the highest intensity value for Fig.~\ref{fig_int_calibration} corresponds to \unit{8$\pm2\times10^{14}$}{W/cm\textsuperscript{2}}.

\section{Scattering detector and spectral filters}
\label{Appendix:scattering_detector_spectral_filters}
Scattering images are recorded in single-shot mode using the front pixel detectors of the LAMP endstation at LCLS~\cite{Strueder2010,Ferguson2015}. The front pixel detector consists of two movable detector panels, each containing 1024x512\,pixel with 75x75\,$\mu$m pixel size. The detectors are covered with a \unit{50}{nm} Al coating to suppress signal from visible light and plasma emission from the target into the extreme ultraviolet spectral regime. For the current experiments with intense \unit{800}{nm} laser pulses, an additional optical filter of \unit{220}{nm} Al on a \unit{0.77}{$\mu$m} LUXfilm polyimide is used and a baffle system is implemented to guide the optical laser pulses through the detector system. The spectral transmission of the filter combined with the detector response~\cite{Strueder2010} result in a net quantum efficiency of 0.48 for detection of the 800\,eV XFEL photons employed for the diffractive imaging with single photon sensitivity.

\section{Laser pulse characterization and timing analysis} \label{Appendix:laser_and_timing}

Our results rely on the accurate determination of the relative delay between the NIR pump pulses and the x-ray scattering probe pulses. The Self Amplification of Spontaneous Emission (SASE) process for Free Electron Lasers as used here produces X-ray pulse shapes whose average temporal shape is well behaved but the individual shots are a near stochastic collection of longitudinal optical modes. When we speak of the ``duration'' of the x-ray pulse, we are therefore speaking of the average duration that is typically a majority fraction of the electron bunch length used to produce the FEL pulse. In our case the electron bunch was measured to correspond to a duration of about 40--50 fs, where the average x-ray pulse duration would be expected in the vicinity of 30--40 femtoseconds FWHM, though we note that the average intensity distribution is not necessarily Gaussian in shape. The NIR pump pulse used in this experiment has been repeatedly measured consistent with 50\,fs duration FWHM at the interaction point. We used the conventional method of optimizing this duration \textit{in situ} by optimizing the ratio of high/low charge states of the strong field ionization of Argon and N$_2$.

\subsection{Calibration of relative delay}
The NIR pump laser and the FEL are independent sources, and as such their relative timing can only be controlled at the accuracy level of the  radio-frequency synchronization system. At the LCLS, this synchronization is typically on the order of a 150\,fs rms and must be corrected by the use of a single-shot x-ray optical cross correlation method~\cite{Schorb:10,Schlotter:10}. The cross correlation method used here is based on the idea of encoding time onto the spectrum of an optical reference~\cite{Bionta:11,MinaRSI}. This so-called ``spectral encoding'' allows for relative measurements with precision below 10\,fs \cite{Harmand13}. 

\begin{figure}
	\centering
	\includegraphics[width=0.7\textwidth]{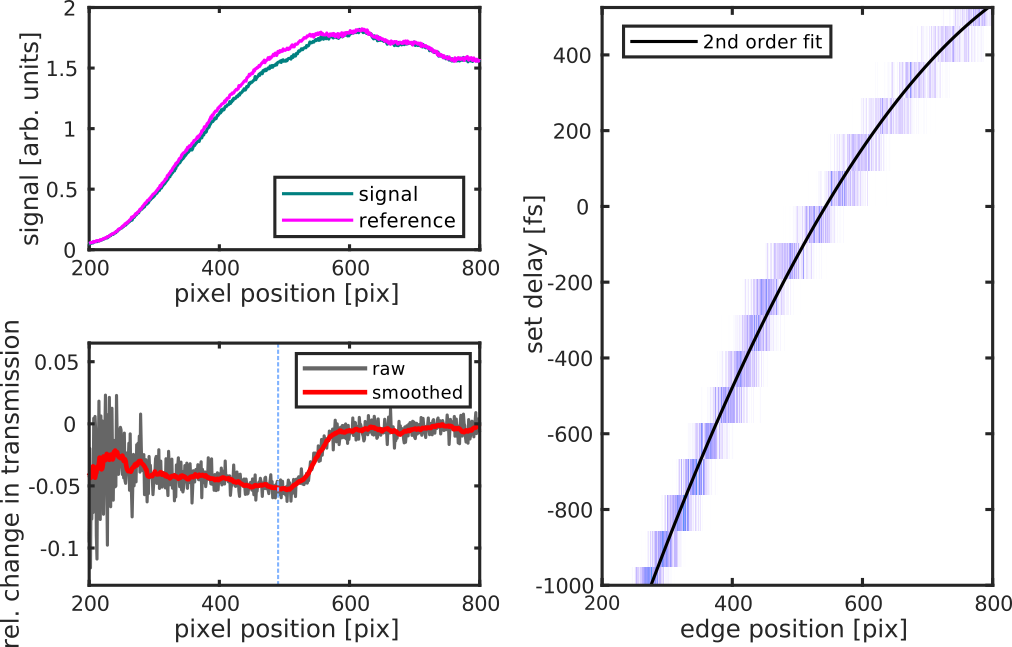}
	\caption{Time tool calibration. (a) Exemplary pair of transmission measurements with (blue) and without X-ray pulse (red). (b) Resulting relative transmission change corresponding to measurements shown in (a) with distinct step. (c) Delay calibration of step position as function of electronic delay via systematic delay scan. The second order polynomial fit is shown as black curve. \label{figS2}}
\end{figure}

The basic observable of the timing tool is the change in the frequency resolved transmission of a super-continuum, generated from a replica of the NIR pulse, through a dielectric plate with (signal) and without (reference) the co-propagating x-ray pulse, see Fig.~\ref{figS2}(a) and (b) for exemplary raw signals and the corresponding relative X-ray induced transmission change, respectively. Due to a spectral chirp of the super-continuum, the spectral transmission change exhibits a step that is associated with the pulse overlap and can be translated into a relative pulse delay. To this end, a systematic scan of the relative pulse delay, e.g. via electronic delay settings, is performed that yields a map for the correlation of the edge position and pulse delay, still including the jitter. A second order polynomial fit to that map provides a mapping function for position and relative delay, see Fig.~\ref{figS2}(c).

\subsection{Finding absolute zero}
In order to identify the absolute time overlap between the NIR and the x-ray pulse, we used the method described in \cite{Glownia:10}
whereby the photo-dissociation of quasi-bound N$_2^{++}$ dications into N$^+$ fragment ions \cite{Coffee06} serves as an indication of the x-ray induced appearance of N$_2^{++}$ via the x-ray core ionization and subsequent normal Auger decay. In particular the delay dependent abundance of the resulting low and high energy cations serves as a measure for the pulse order. Due to the high ionization potential of $N_2$, the attenuated NIR-pulse is unable to excite or ionize the neutral molecule, leaving the ionization by the subsequent X-ray pulse and the resulting population of quasi-bound $N_2^{++}$ almost unchanged. However, initial irradiation of the neutral molecules by the X-ray  generates a population of quasi-bound $N_2^{++}$ that, after further excitation by the NIR pulse, results in dissociation into a pair of $N^+$ with corresponding recoil energy due to Coulomb repulsion. 
\begin{figure*}
	\centering
	\includegraphics[width=1.0\textwidth]{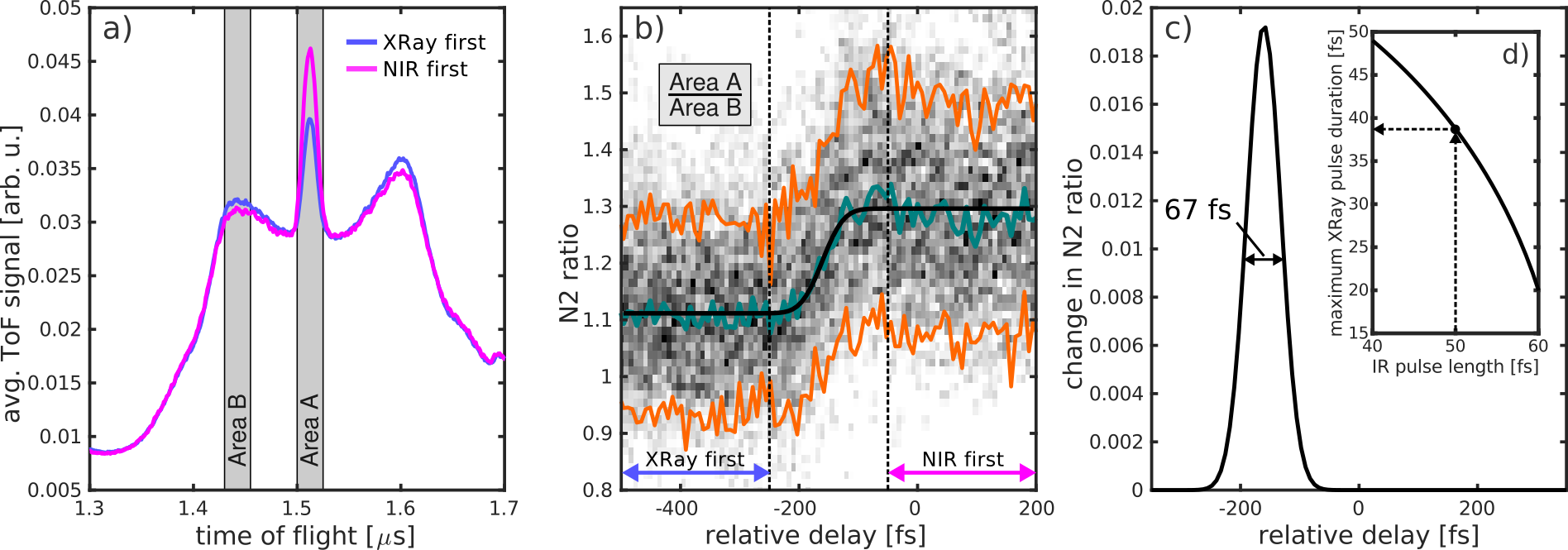}
	\caption{(a) Averaged time-of-flight spectra containing  $N^+$ and $N_2^{++}$ ion signals for shots with X-ray first and NIR first excitation [delay ranges are indicated in (b)]. The highlighted areas define the integration regions for the respective yields of $N_2^{++}$ (area A) and $N_2^+$ ions with recoil (area B). (b) 2D-histogram of single-shot ion yield ratios $A/B$ with center and width of corresponding Gaussian fits indicated by cyan and orange lines, respectively. (c) Rate of change of the delay-dependent (fitted) ratio center and resulting X-ray pulse duration as function of the NIR pulse duration (see inset). The arrows indicate the X-ray pulse duration resulting for the measured NIR pulse duration of $50\,$fs. \label{fig_a_find_zero}}
\end{figure*}

Fig.~\ref{fig_a_find_zero}(a) presents averaged time-of-flight spectra for shots with the X-ray pulse first (blue) or the NIR pulse first (purple). Note that area A indicates the recoil-free signal from $N_2^{++}$ that exhibits higher yield if the NIR arrives first. Area B is representative for $N^+$ signal with recoil and shows increased yield when the NIR is late. 
The ratio of the signals integrated within   these areas is used to trace the pulse order as function of the (not offset corrected) delay. Fig.~\ref{fig_a_find_zero}(b) shows the distribution of the single-shot yield ratios from a systematic delay scan. For each delay bin, the distribution can be described well by a Gaussian envelope. The resulting delay-dependent center position (width) of the Gaussian fits is shown as cyan (orange) curve. The step around \unit{-160}{fs} is indicative of pulse overlap, i.e. zero delay.

Assuming no-depletion and linear response for both the X-ray and NIR excitations processes, the derivative of the delay-dependent ratio signal directly reflects the cross correlation of the pulses envelopes 
\begin{eqnarray}
    \frac{d}{d\tau} \frac{A(\tau)}{B(\tau)}\propto (I_{\rm xray}* I_{\rm NIR})(\tau).
\end{eqnarray}

A fit of the step feature in the evolution of the delay-dependent centers in Fig.~\ref{fig_a_find_zero}(b) with an error function (black) and taking the derivative of the latter yields the cross-correlation in Fig.~\ref{fig_a_find_zero}(c), whose peak is located at $\Delta t_0=$\unit{-160}{fs}, defining zero delay. Assuming Guassian pulses the FWHM-width of the cross correlation of \unit{67}{fs} can be used to estimate the mean X-ray pulse duration to about 38\,fs from the knowledge of the NIR pulse duration of 50\,fs (see inset). 

\section{Small angle scattering of ellipsoids in the weak scattering limit} \label{Appendix:small_angle_scattering}
\subsection{Evaluation of the scattering integral in Born approximation}
To derive the formalism used to describe the scattering of ellipsoids in our study we depart from the general solution to the classical scattering problem in the limit of weak scattering, i.e. neglecting multiple scattering (Born approximation). Assuming a linearly polarized incident electric field $\vec{E}_{inc}=\vec{E}_{0}e^{-i \vec{k}_{in}\vec{r}}$ with wave vector $\vec{k}_{\rm in}$ and an arbitrarily shaped scattering object defined via the scattering electron density $n(\vec{r})$ we can calculate the scattered field amplitude from
\begin{eqnarray}
E(\vec{q})&=E_0 \frac{r_e}{l_{det}}\int\int\int n(\vec{r}) e^{-i \vec{q}\cdot\vec{r}} \mathrm{d}^3r . \label{eq:scatt_kinematic} 
\end{eqnarray}
with momentum transfer vector $\vec{q}$ defined as the difference between incoming and outgoing wave vector according to $\vec{q}=\vec{k}_{\rm out}-\vec{k}_{\rm in}$, classical electron radius $r_e$ and detector distance  $l_{det}$. {Note that the angular dependence of the scattered field amplitude due to polarization effects can be neglected for small scattering angles.}

For the following discussion we consider, without loss of generality, an incident wave vector along the z-axis $\vec{k}_{in} = k\vec{e}_z$. We further limit the discussion to a small angle scattering scenario with $|q|\ll |k_{\rm in}|$) such that we can now assume $q_z\simeq0$ and rewrite Eq.~\ref{eq:scatt_kinematic} as follows:
\begin{eqnarray}
E(\vec{q})&=E_0 \frac{r_e}{l_{det}}\int_{-\infty}^{\infty}\int_{-\infty}^{\infty}\int_{-\infty}^{\infty} n(\vec{r}) e^{-i ({q_x x +q_y y+q_z z})} \mathrm{d}x\mathrm{d}y\mathrm{d}z \nonumber\\ 
&=E_0 \frac{r_e}{l_{det}}\int_{-\infty}^{\infty}\int_{-\infty}^{\infty}\underbrace{\int_{-\infty}^{\infty} n(\vec{r})\mathrm{d}z}_{n_{2d}(x,y)} e^{-i ({q_x x +q_y y})} \mathrm{d}x\mathrm{d}y \nonumber\\
&=E_0 \frac{r_e}{l_{det}}\int_{-\infty}^{\infty}\int_{-\infty}^{\infty}n_{2d}(x,y) e^{-i ({q_x x +q_y y})} \mathrm{d}x\mathrm{d}y \nonumber\\
&=E_0 \frac{r_e}{l_{det}} \mathcal{F} \left[n_{2d}\right](q_x,q_y) \label{eq:scatt_kinematic_2dFourier},
\end{eqnarray}
which reflects that the scattering intensity in the small angle scattering limit is determined by the Fourier transform of the projected density $n_{2d}$.

If we are only interested in a line profile instead of the full 2d scattering image, e.g. for $q_y=0$, we can further simplify the expression and write:
\begin{eqnarray}
E(q_x,q_y=0)&=E_0 \frac{r_e}{l_{det}}\int_{-\infty}^{\infty}\int_{-\infty}^{\infty}n_{2d}(x,y) e^{-i ({q_x x +q_y y})} \mathrm{d}x\mathrm{d}y \nonumber\\ 
&=E_0 \frac{r_e}{l_{det}}\int_{-\infty}^{\infty}\underbrace{\int_{-\infty}^{\infty}n_{2d}(x,y)\mathrm{d}y}_{n_{1d}(x)} e^{-i ({q_x x})} \mathrm{d}x \nonumber\\
&=E_0 \frac{r_e}{l_{det}}\int_{-\infty}^{\infty}n_{1d}(x) e^{-i ({q_x x})} \mathrm{d}x \nonumber\\
&=E_0 \frac{r_e}{l_{det}}\mathcal{F}\left[n_{1d}\right](q_x)\label{eq:scatt_kinematic_small_angle_1d}.
\end{eqnarray}
which relates the scattered field along a specific polar angle with the Fourier transform of the density projected onto the corresponding axis. Based on the above equation we will now derive solution strategies for the density distributions that are most relevant to our work, i.e. that of homogeneous spheres and ellipsoids with sharp and soft surfaces.

\subsection{Analytic solution for a homogeneous sphere with a sharp surface}
As the first step we will revisit the well-known solution for a homogeneous sphere with a sharp surface described by the density profile
\begin{eqnarray*}
n(r)=\left \{\begin{array}{c c} 0 & \quad r> R \\ n_0 &  \quad r\le R \end{array}\right.
\end{eqnarray*}
with the homogeneous electron density $n_0$. Due to the radial symmetry we only need to calculate the scattering for a single scattering direction, i.e. we are only interested in the double projected density $n_{1d}(x)$. The corresponding integral can be calculated in polar coordinates
\begin{eqnarray*}
n_{1d}(x)=n_0 \int_{0}^{2\pi}\int_{0}^{\rho_{\rm max}}  \rho\rmd\rho\rmd\varphi =n_0 \pi \rho_{\rm max}^2
\end{eqnarray*}
where $\rho_{max}(x)$ describes the radial extension of the sphere for the given position on the projection axis $x$ with
\begin{eqnarray*}
\rho_{max}(x)=\left \{\begin{array}{c c} \sqrt{R^2-x^2} & \quad x\le R \\ 0 &  \quad x>R \end{array}\right.
\end{eqnarray*}
Inserting this result yields
\begin{eqnarray} \label{eq:n1d_sphere_sharp}
n_{1d}(x)=\left \{\begin{array}{c c} 0 & \quad x> R \\ n_0 \pi(R^2-x^2) &  \quad x\le R \end{array}\right. 
\end{eqnarray}
and allows to obtain the scattered field by solving the corresponding Fourier integral
\begin{eqnarray*}
E(q_x)&=E_0 \frac{r_e}{l_{det}}\int_{-\infty}^{\infty}n_{1d}(x) e^{-i ({q_x x})} \mathrm{d}x \nonumber\\
&=n_0 E_0 \frac{r_e}{l_{det}}\int_{-R}^{R}\pi(R^2-x^2) e^{-i ({q_x x})} \mathrm{d}x \nonumber\\
&=E_0 \frac{r_e}{l_{det}} n_0 4 \pi R^3 \left(\frac{\sin{(q_x R)}- q_x R \cos{(q_x R)}}{q_x^3 R^3}\right) \nonumber\\
\end{eqnarray*}
It is often beneficial to work with the normalized scattered field $E_n$ ($E_n(0)=1$) which reads
\begin{eqnarray*}
E_n(q)&=3\left(\frac{\sin{(q R)}- q R \cos{(q R)}}{q^3 R^3}\right).
\end{eqnarray*}
Note, that we can drop the subscript $x$ of the momentum transfer vector due to the radial symmetry.

\subsection{Analytic solution for a homogeneous ellipsoid with a sharp surface} \label{sec:hom_ell_sharp}
The doubly-projected density $n_{1D}$ that is needed to evaluate the scattering integral in Eq.~\ref{eq:scatt_kinematic_small_angle_1d} for an object is also an important quantity in the field of reconstruction algorithms in computer tomography (CT), where it is known as the Radon transform. In particular, the Radon transform of arbitrarily shaped and oriented ellipsoids are of wider interest and are analytically known \cite{zhu_ellipsoid}
\begin{eqnarray}
n_{1D}(\rho,\vec{n})=\left \{\begin{array}{c c} 0 & \quad |\rho| \ge \zeta \\ n_0 \pi \frac{abc}{\zeta^3}\left(\zeta^2-\rho^2\right) &  \quad |\rho|<\zeta \end{array}\right.
\end{eqnarray}
where $\vec{n}=(\sin{\theta\cos{\varphi},\sin{\theta}\sin{\varphi},\cos{\theta}})^t$ is the normalized vector characterizing the direction the density is projected on and $\rho=\vec{r}\vec{n}$ is the remaining coordinate along this direction. The parameters $a$, $b$ and $c$ determine the ellipsoids length along the x, y and z direction for an unrotated ellipsoid. The effects of the dimension and orientation of the  ellipsoid are now encapsulated in the factor $\zeta$:
\begin{eqnarray*}
\zeta^2=&a^2(\cos{\theta}\sin{\psi_2}\sin{\psi_3}\\
&+\sin{\theta}\sin{\varphi}(\cos{\psi_3}\sin{\psi_1}+\cos{\psi_1}\cos{\psi_2}\sin{\psi_3})\\
&+\cos{\varphi}\sin{\theta}(\cos{\psi_1}\cos{\psi_3-\cos{\psi_2}\sin{\psi_1}\sin{\psi_3}}))^2\\
+&b^2(\cos{\theta}\sin{\psi_3}\cos{\psi_3}\\
&-\sin{\theta}\cos{\varphi}(\cos{\psi_1}\sin{\psi_3}+\sin{\psi_1}\cos{\psi_2}\cos{\psi_3})\\
&+\sin{\varphi}\sin{\theta}(\cos{\psi_1}\cos{\psi_2}\cos{\psi_3}-\sin{\psi_1}\sin{\psi_3}))^2\\
+&c^2(\cos{\theta}\cos{\psi_2}-\sin{\theta}\sin{(\varphi-\psi_1)}\sin{\psi_2})^2
\end{eqnarray*}
which gives us the effective dimension of the projected ellipsoid along the direction of the projection vector $\vec{n}$. The angles $\psi_1, \psi_2$ and $\psi_3$ are Euler angles in x-convention for the rotation of the ellipsoid (rotation first by angle $\psi_1$ about the z-axis, secondly by angle $0\ge\psi_2\le\pi$ about the x-axis, and finally by angle $\psi_3$ about the z-axis again). For our purposes we can, without loss of generality, assume a detector in the x-y plane (our propagation direction is z), i.e. $\theta=\pi/2$ such that the angle $\varphi$ is essentially the polar angle defining the line-out from the 2d scattering pattern. Doing this already simplifies the above expressions considerably:
\begin{eqnarray*}
\zeta^2=&a^2(\sin{\varphi}(\cos{\psi_3}\sin{\psi_1}+\cos{\psi_1}\cos{\psi_2}\sin{\psi_3})\\
&+\cos{\varphi}(\cos{\psi_1}\cos{\psi_3-\cos{\psi_2}\sin{\psi_1}\sin{\psi_3}}))^2\\
+&b^2(-\cos{\varphi}(\cos{\psi_1}\sin{\psi_3}+\sin{\psi_1}\cos{\psi_2}\cos{\psi_3})\\
&+\sin{\varphi}(\cos{\psi_1}\cos{\psi_2}\cos{\psi_3}-\sin{\psi_1}\sin{\psi_3}))^2\\
+&c^2(-\sin{(\varphi-\psi_1)}\sin{\psi_2})^2 .
\end{eqnarray*}
For any given projection direction $\vec{n}$ the expression for the double projected density is structure-wise identical to that for a homogeneous sphere, c.f. Eq.~\ref{eq:n1d_sphere_sharp}. However, the value of the radius equivalent $\zeta$ changes as a function of the projection direction. Before we discuss the implications of this aspect, we further reduce the complexity of the expression for $\zeta$ with the help of trigonometrical identities such that
\begin{eqnarray*}
\zeta^2=&a^2(\sin{\varphi'}\cos{\psi_2}\sin{\psi_3}+\cos{\varphi'}\cos{\psi_3})^2\\
+&b^2(\sin{\varphi'}\cos{\psi_2}\cos{\psi_3}-\cos{\varphi'}\sin{\psi_3})^2\\
+&c^2(\sin{\varphi'}\sin{\psi_2})^2,
\end{eqnarray*}
where $\varphi'=\varphi-\psi_1$. It can be shown, that for any given set of parameters $a$, $b$, $c$, $\psi_1$, $\psi_2$, $\psi_3$ and $\varphi$ the description of $\zeta^2$ can be reduced to an even simpler functional form:
\begin{eqnarray*}
\zeta^2=&a_{eff}^2 \cos{(\varphi'+\Delta)}^2\\
+&b_{eff}^2 \sin{(\varphi'+\Delta)}^2
\end{eqnarray*}
where the effective parameters $a_{eff}$ and $b_{eff}$ describe the dimensions of the projection and $\Delta$ describes its orientation in the x-y plane. 

Now we can solve the Fourier integral for each projection direction individually in full analogy to the spherical case discussed above. The result is a scattered field that resembles the scattering of a sphere for any given direction $\varphi''=\varphi'+\Delta$, but with a direction dependent radius $R(\varphi'') \equiv \zeta$. The result for the scattered field amplitude reads
\begin{eqnarray*}
&E(q,\varphi'') = \\
&E_0 \frac{r_e}{l_{det}} n_0 4 \pi abc  \left(\frac{{\sin(q R(\varphi''))}- q R(\varphi'') \cos{(q R(\varphi''))}}{q^3 R(\varphi'')^3}\right) \nonumber\\
\end{eqnarray*}
The corresponding normalized fields ($E_n(0,\varphi'')=1$) can be written as
\begin{eqnarray} 
E_n(q,\varphi')&=3\left(\frac{{\sin(q R(\varphi''))}- q R(\varphi'') \cos{(q R(\varphi''))}}{q^3 R(\varphi'')^3}\right). \label{eq:norm_field_ell_sharp}
\end{eqnarray}
which resembles the solution of a sharp sphere, but with an angular dependent radius.

\subsection{Solution for a sphere with a soft surface}
Next, we will consider spherical objects with a soft surface density profile, akin to the one used in the main manuscript, i.e.
\begin{eqnarray*}
n(r)=\frac{n_0}{\left[\exp(\frac{r-R}{\lambda \sigma})+1\right]^\sigma}
\end{eqnarray*}
where $n_0$ is the core density, $R$ is the core radius, $\sigma$ is the edge softness parameter, and $\lambda$ is the scale length of the density decay. Again, due to the radial symmetry we only need to calculate the scattering for a single scattering direction, i.e. we are only interested in the double projected density $n_{1d}(x)$. The corresponding integral can be calculated in polar coordinates
\begin{eqnarray*}
n_{1d}(x)=\int_{0}^{2\pi}\int_{0}^{\infty} n(r)  \rho \, \rmd\rho\rmd\varphi \\
=2 \pi \int_{0}^{\infty} n(\sqrt{x^2+\rho^2})  \rho \, \rmd\rho \\
\end{eqnarray*}
In contrast to the scenarios with a sharp surface discussed above, this integral can in general not be solved analytically. However, since only a single one-dimensional integral has to be evaluated to calculate the 2D scattering pattern, a direct numerical integration is a fast and viable approach.

\subsection{Approximation for an ellipsoid with a soft surface}
Eventually we are looking for an efficient way to compute the scattering pattern for an arbitrarily dimensioned and oriented ellipsoid with a softened density profile given by
\begin{equation}
n(\vec{r})= \frac{n_0}{\left[\exp(\frac{\kappa(\vec{r})}{\lambda \sigma})+1\right]^\sigma}
\label{eq_profile_appendix},
\end{equation}
where $n_0$ is the core density, $\kappa$ is the local surface offset of the sampling point $\vec{r}$ (i.e. its distance to the closest point on the ellipsoidal control surface, see main text), $\sigma$ is the edge softness parameter, and $\lambda$ is the scale length of the density decay. For this type of profile no general solutions for the singly- or doubly-projected densities $n_{2D}$ and $n_{1D}$ are known. An apparent alternative way to obtain these projections involves the three-dimensional sampling of the density on a grid and subsequent numerical projection. This requires the evaluation of the perpendicular distance from the control ellipsoid surface $\kappa$ for all given sampling points. In the limit of a sphere this turns into the radial distance from the surface, i,e, $\kappa=r-R$, which is easily calculated. However, in the general case of an ellipsoid, the evaluation of the perpendicular (or closest) distance is not trivial. In fact, its computation involves numerical root finding procedures~\cite{eberly_ell_dist}, which practically precludes application of this method in forward fitting approaches. 

\begin{figure*}[t]
\centering
\includegraphics[width=1.0\textwidth]{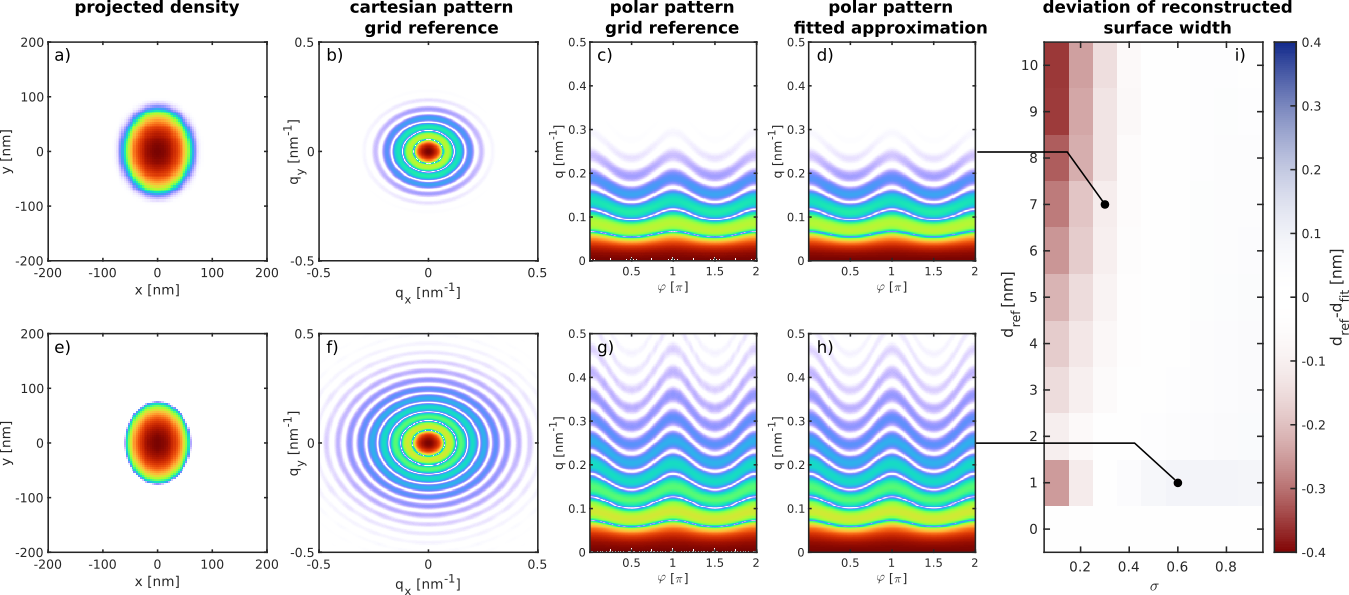}
\caption{Systematic study of the quality of our approximation for a softened ellipsoid with semi-axis \unit{60}{nm}, \unit{80}{nm} and \unit{60}{nm} rotated 45 degree about the x-axis. The scale length $\lambda$ and the edge softness $\sigma$ have been varied systematically within the ranges most relevant for this work. The density has been sampled directly on a grid and subsequently projected to obtain $n_{2D}$ and the reference 2D scattering pattern. The corresponding polar representation has then been fitted using the described approximation (with fixed edge softness $\sigma$). The projected densities, reference cartesian- and polar scattering images as well as the fitted scattering pattern are shown for two examplary configurations in panels (a) to (h).  Panel (i) shows the difference between the actual scale length of the reference $d_{ref}$ and the reconstructed scale length from the fit $d_{fit}$. \label{fig:approximation_quality}}
\end{figure*}

To solve this problem, we propose an approximation that is based on a combination of the solution for a homogeneous ellipsoid with sharp surface and the solution for a soft sphere. To recall, for an ellipsoid with sharp surface the scattering profile for any given scattering direction is given by that of sharp sphere, the ellipsoidal character enters the description in the form of an angular dependent effective sphere radius. Here, for the ellipsoid with a soft surface, we use a similar approach by replacing the solution of a sharp sphere with that of a soft sphere for a given scattering direction. This approximation will work best for small eccentricities and even becomes exact for a sphere. A systematic analysis of the approximation quality is shown in figure \ref{fig:approximation_quality}. Note that the eccentricity of the particles used in Fig.~\ref{fig:approximation_quality} has been chosen larger than the average eccentricity of the particle in the experiment to show the outstanding quality of the approximation in the considered parameter regime.

\subsection{Forward fitting procedure} \label{Appendix:fitting_procedure}
The main tool for the fitting procedure will be the above approximation of the scattering pattern of a soft ellipsoid by that of a soft sphere with angle dependent effective radius. In a first step the recorded experimental scattering pattern is transformed into a polar representation. Polar coordinates are the natural coordinates of the numerical scheme for ellipsoid scattering described above and are therefore optimally suited for a comparison of experiment and theory. Furthermore, for properly chosen resolution this step intrinsically applies a sub-sampling of the scattering signal at large scattering angles where the signal strength and therefore the signal-to-noise ratio naturally decreases.

In our approach a simulated scattering image is fully defined by a set of 7 parameters. Besides the parameters defined above, i.e. the density projection dimensions $a_{\rm eff}$ and $b_{\rm eff}$, the surface scale length $\lambda$, the projections orientation $\Delta$ and the edge softness $\sigma$ that are necessary to compute the normalized scattering pattern $I_n$, we consider the incident field intensity $I_0$ and a constant background signal $I_{bg}$.
\begin{eqnarray}
    I = I_0 I_n(a_{\rm eff},b_{\rm eff},\lambda,\Delta,\sigma)+I_{\rm bg}
\end{eqnarray}

\begin{figure*}[t]
	\centering
	\includegraphics[width=1.0\textwidth]{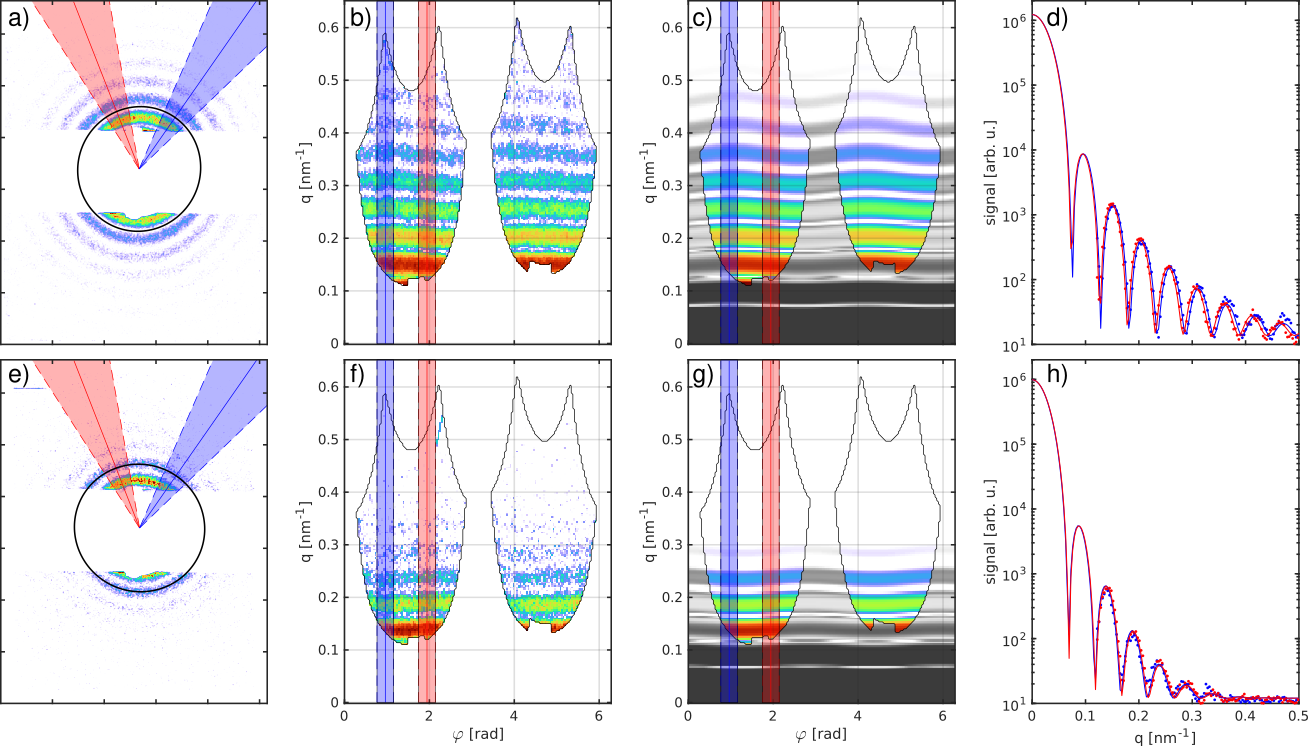}
	\caption{Comparison of recorded scattering images and corresponding fits for representative pattern with sharp surface (top panels) and soft surface (bottom panels). (a,e) Cartesian representation of recorded scattering pattern. (b,f) Polar representation of recorded scattering pattern. The active areas associated with the two detector panels is outlined by the solid black line. (c,g) Polar representation of corresponding fits. Parts relevant for the comparison with the experiment are color-coded. (d,h) Comparison of line-outs of recorded and fitted pattern at indicated scattering angles (see red and blue shaded areas in other panels). \label{fig:supp_expamples}}
\end{figure*}

In our forward fitting procedure parameter optimization has been achieved using the ``trust-region-reflective'' algorithm embedded in Matlabs optimization toolbox. Figure \ref{fig:supp_expamples} shows two representative exemplary optimization results in comparison with the experimental reference pattern for a sharp surface (top panels) and a soft surface (bottom panels).

\section*{References}
\bibliographystyle{iopart-num}
\bibliography{Peltz_references}

\end{document}